\documentclass[aps,prl,twocolumn,groupedaddress,10pt]{revtex4}
\usepackage{graphicx}  
\usepackage{subfigure}
\usepackage{color}

\newcommand{\bq}{\begin{equation}}
\newcommand{\eq}{\end{equation}}
\newcommand{\ba}{\begin{eqnarray}}
\newcommand{\ea}{\end{eqnarray}}
\newcommand{\dd}{{\rm d}}

\begin{document}
\title{A physics-explicit model of bacterial conjugation shows the stabilizing role of the conjugative junction}
\author{Jakub Pastuszak$^{1,2}$, Bartlomiej Waclaw$^{1,3}$}
\affiliation{$^1$School of Physics and Astronomy, The University of Edinburgh, Peter Guthrie Tait Road, Edinburgh EH9 3FD, United Kingdom\\
$^2$currently at Orbium International, Al. Jerozolimskie 98, 00-807 Warsaw, Poland\\
$^3$Centre for Synthetic and Systems Biology, Edinburgh EH9 3FD, United Kingdom}

\begin{abstract}
Conjugation is a process in which bacteria exchange DNA through a physical connection (conjugative junction) between mating cells. Despite its significance for processes such as the spread of antibiotic resistance, the role of physical forces in conjugation is poorly understood. Here we use computer models to show that the conjugative junction not only serves as a link to transfer the DNA but it also mechanically stabilises the mating pair which significantly increases the conjugation rate. We discuss the importance of our findings for biological evolution and suggest experiments to validate them.
\end{abstract}


\maketitle

Bacterial DNA evolves by two mechanisms: spontaneous alterations (mutations) which occur mostly during replication, and gene transfer between unrelated cells. The latter process is usually called horizontal gene transfer (HGT) \cite{syvanen2001horizontal} since it can be pictured as horizontal links in phylogenetic trees which represent the evolution of organisms over time (vertical direction). 
HGT is a powerful force in biological evolution and is believed to drive processes such as the acquisition of antibiotic resistance genes \cite{Hiby2010,Davison19pp} and virulence \cite{ochman_lateral_2000}. Since HGT involves direct interactions between mating cells, its dynamics is very rich and has been extensively studied by physicists \cite{park_phase_2007,he_spontaneous_2009,chia_statistical_2011,deem_statistical_2013,court_parasites_2013,venegas-ortiz_speed_2014,freese_genetic_2014}.

One of the mechanisms of HGT is bacterial conjugation \cite{lederberg1946gene,sorensen_studying_2005} in which a plasmid - a small DNA molecule independent from the main bacterial chromosome - is copied from a donor cell to a recipient cell. Conjugation involves an initiation step upon contact between the donor and the recipient, the formation of a junction between the cells, and the transfer of a copy of the plasmid to the recipient cell. However, the details of these processes remain elusive. In particular, it is unclear what type of mechanical connection between conjugating cells is required for plasmid transfer \cite{barocchi_bacterial_2013,samuels2000conjugative}, how shear forces due to cellular movement affect conjugation \cite{achtman_cell-cell_1978}, and whether the sexual pilus - a thin tube used during the initiation phase to bring the two cells together - is also involved in transferring the plasmid \cite{barocchi_bacterial_2013}.

In this work we develop a computational framework to study conjugation in bacterial colonies in which cells are in close mechanical contact. This is the case of naturally occurring cellular agglomerates such as biofilms \cite{donlan2002biofilms} and laboratory-cultured bacterial colonies \cite{ben-jacob_generic_1994}. Previous models of conjugation often assume well-mixed, unstructured populations \cite{levin1979kinetics,levin1980population,simonsen1990estimating,duncan1995fitnesses,park_phase_2007,he_spontaneous_2009,chia_statistical_2011,deem_statistical_2013} which is appropriate for bacteria cultured in shaken flasks or chemostats, but not for bacterial colonies growing on surfaces. Existing spatial models of conjugation \cite{court_parasites_2013,venegas-ortiz_speed_2014,freese_genetic_2014} do not include mechanical interactions between cells, and conjugation is represented as an idealized, one-step stochastic process. 

\begin{figure}[b]
\includegraphics*[width=\columnwidth, bb=50 60 550 390]{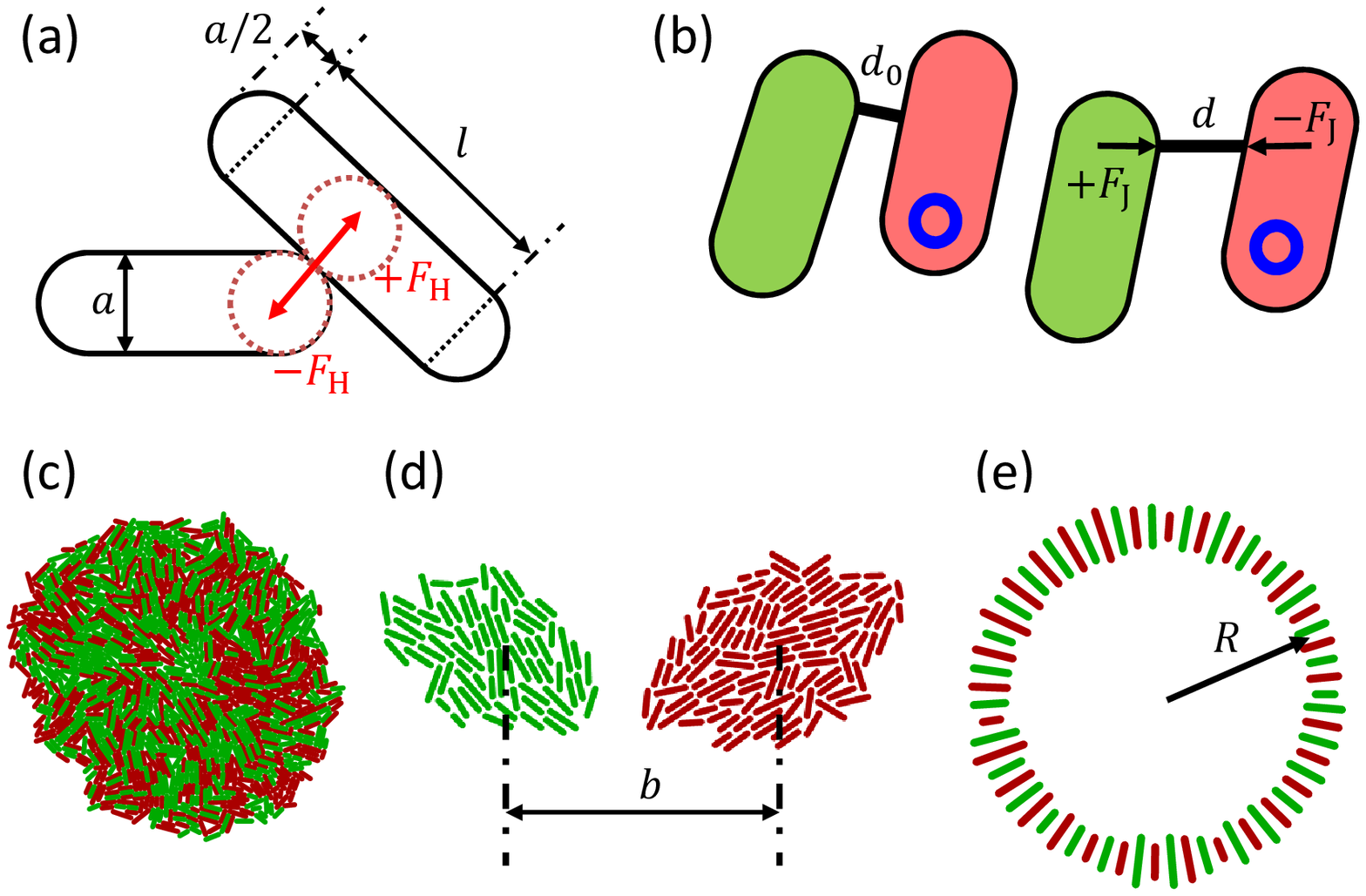}
\caption{The computer model. (a) Cells are modelled as spherocylinders of length $l$ and diameter $a$, interacting through Hertzian repulsive forces $F_H$. (b) Conjugative junction in Models 2 and 3 is an elastic spring of rest length $d_0$ and spring constant $k$. The spring creates a restoring force $F_J$ when the distance $d$ between the conjugating cells (donor = red cell with a blue circle representing the plasmid, recipient = green cell) is different to $d_0$. (c) A snapshot of an exponentially growing colony with a 1:1 mixture donors:acceptors. (d) Colliding colonies. $b$ is the distance between the two cells that initiated the colonies. (e) Initial configuration of cells in the linearly-expanding colony simulation. }
\end{figure}

Here we take a physics-oriented approach and explicitly model mechanical interactions such as the repulsion between cells, friction between cells and the surface on which they live, and cell-cell adhesion caused by the formation of conjugative junctions. 
We investigate how these interactions affect conjugation in three scenarios motivated by typical laboratory experiments: an exponentially-growing colony starting from a mixture of donor and recipient cells, a colony of donors colliding with a colony of recipients, and a mixed colony with a growing layer of cells and a quiescent core. 
We show that if the junction exerts no force, the apparent conjugation rate is very low in all these scenarios due to relative movement of donors and acceptors, but it increases significantly if the junction is strong enough to oppose the motion.

To model a colony of bacterial cells we use the algorithm described in Ref. \cite{farrell_mechanically_2013}. Cells are modelled as non-motile spherocylinders with variable length $l=2\dots 4\mu$m (excluding the spherical caps) and fixed diameter $a=1\mu$m (Fig. 1a). The movement of cells is restricted to the 2d $xy$ plane. Cells consume nutrients that diffuse in the $xy$ plane, elongate, and divide when they reach a maximal length $\approx 4\mu$m (see Appendix A for details). The repulsion force between cells is modelled as an elastic, Hertzian-like contact force $F_{\rm H}=Ea^{1/2}h^{3/2}$, where $E$ is proportional to the effective elastic modulus of cells (we assume $E=100$kPa), and $h=a-d$ where $d$ is the minimal distance between the cells ($h=0$ when the cells just touch). The movement of cells is simulated using Newton's overdamped equations of motion (Appendix A).

Conjugation is implemented as follows.
Each cell is either a donor (D), a recipient (R) or a transconjugant (T). Recipient cells do not have the conjugative plasmid. Donor cells are the cells that had the plasmid at the beginning of the simulation ($t=0$), as well as descendants of such cells. Transconjugants are recipient cells that received the plasmid either from the donors or other transconjugant cells. All cells are assumed to grow at the same rate $g$, i.e., there is no fitness cost or benefit associated with the plasmid. We also neglect segregative plasmid loss \cite{krone2007modelling,de2007stability} for simplicity.

We consider three different models of conjugation. In Model 1, if R is in direct mechanical contact ($h>0$) with either D or T, and both mating cells are growing ($f(c)>0$), R transforms into T (i.e. it acquires the plasmid) with rate $\mu$ $[h^{-1}]$. Plasmid transfer is instantaneous in this case; once initiated (with rate $\mu$) it always succeeds. The condition $f(c)>0$ ensures that only cells that have sufficient nutrients engage in conjugation \cite{simonsen1990estimating}.

In Model 2 (Fig. 1b), cells form a conjugative junction with rate $\mu$, but plasmid transfer requires that the cells remain near each other for $0.1$h ($6$ min). Newly formed transconjugant cells can conjugate again only after a lag period of $7$ min. The assumed values of the conjugation and lag times are typical; experimentally observed values range from a few mins to over 10 mins \cite{michel-briand_inhibition_1985,andrup1998kinetics,andrup1999comparison}.
We assume that the junction can form only if the two interacting bacteria are less than $100$nm apart; experiments \cite{samuels2000conjugative} indicate that the distance of closest approach during conjugation is $10\dots 40$nm but it is likely that the minimum distance required to initiate conjugation is slightly larger \cite{barocchi_bacterial_2013}. The length $d_0$ of the junction and the position of the two ends relative to the position of the conjugating cells at the time when it is formed is remembered, and if the distance $d$ between the cells increases such that $\epsilon=(d-d_0)/d_0$ is larger than some critical strain $\epsilon_0$, the junction breaks and conjugation is interrupted. 

Model 3 is the same as model 2 except that junctions exert force $\vec{F}_J=-k (d-d_0) \vec{n}$ where $\vec{n}$ is a normalized vector pointing in the direction of the junction and $k$ is the spring constant of the junction (Fig. 1b). The force helps to stabilize the mating pair of cells, thus potentially prolonging the phase of direct mechanical contact.

{\it Exponentially-expanding colony.}
We first consider a scenario in which the colony is initiated from a mixture of donors and recipients and is allowed to grow exponentially in time (Fig. 1c). 
Nutrients are unlimited and all cells grow with the same rate. The initial colony is prepared by placing a single cell at the origin and let it replicate until there is $N_0=250$ cells. The cells are then randomly assigned two types (donors and recipients) to create a 1:1 mixture of D:R, and growth is resumed. 
The simulation is stopped when the number of cells reaches $N=2^3\times 250=2000$ which corresponds to 3 generations. 
This mimics a typical conjugation experiment \cite{del2012determination,reisner_situ_2012} in which a mixture of cells is placed on the surface of nutrient-infused agarose and allowed to grow and mate for a fixed time. 
We are interested in the fraction $x$ of transconjugants in the population of cells at the end of the simulation. $x$ can be determined experimentally \cite{lilley2002transfer,del2012determination} and is a convenient measure of conjugation efficiency. 

We begin by investigating the instantaneous plasmid-transfer Model 1. Figure 2a shows the proportion $x$ of transconjugants in the colony for different $\mu$. As expected, $x$ increases monotonically with $\mu$ and reaches a plateau $x\approx 1/2$ for large $\mu$. In this simple model, the spatial distribution of cells plays only a minor role. Indeed, $x(\mu)$ follows the curve predicted by a well-mixed model of conjugating cells, modulo a correction that accounts for the reduced number of mating pairs due to steric hindrance. Let $n_D,n_R,n_T$ be the number of donor, recipient, and transconjugant cells. If we assume that all cells replicate at the same rate $g$ and that the rate of conjugation (a two-body process) is proportional to the product of the density of recipients and donors+transconjugants, we have
\ba
	\dd n_D/\dd t &=& g n_D, \label{eq:nd} \\
	\dd n_T/\dd t &=& g n_T + \mu q g n_D(n_T+n_D)/n, \\
	\dd n_R/\dd t &=& g n_R - \mu q g n_D(n_T+n_D)/n, \label{eq:nr}
\ea
where $n=n_D+n_R+n_T$ and $q$ may be interpreted as the effective number of neighbours that are available for conjugation. Solving Eqs. (\ref{eq:nd}-\ref{eq:nr}) with the initial condition $n_T(0)=0, n_D(0)=N_0/2, n_R(0)=N_0/2$ gives
\bq
	x = \frac{n_T}{n} = \frac{\left(\frac{N}{N_0}\right)^{\mu q}-1}{2 \left[\left(\frac{N}{N_0}\right)^{\mu  q}+1\right]} . \label{eq:xwm}
\eq
Figure 2a shows that Eq. (\ref{eq:xwm}) fits the simulation very well when $q=7$.

We now turn to Model 2 in which cells must remain in contact for some time for conjugation to occur, but the junction does not exert any force on the cells. Figure 2b shows the fraction $x$ of transconjugants for different values of critical strain $\epsilon_0$ for which the junction breaks. We see that for all but the highest value of $\epsilon_0$, the maximum fraction $x$ is significantly lower than the maximum possible value $1/2$. This shows that, for large initiation rates $\mu$, conjugation initiation is no longer the limiting factor. Instead, the time the two mating cells spend in  proximity becomes the limiting factor. The larger the critical strain, the more the cells can move during conjugation before they separate and conjugation is interrupted. This is supported by Fig. 2d in which we plot $x$ versus $\epsilon_0$ for $\mu=0.016$h$^{-1}$.

We finally investigate Model 3 ($k>0$). To estimate the value of $k$, we assume that the conjugative junction is a part of the cell envelope \cite{samuels2000conjugative} and is a tube of diameter $d_{\rm out}=100$nm, length $l_{\rm junction}=100$nm \cite{samuels2000conjugative}, and wall thickness $\Delta d=4$nm \cite{tuson_measuring_2012}. The spring constant is then $k=A_{\rm junction}E_{\rm wall}/l_{\rm junction}$ where $A_{\rm junction} = \pi d_{\rm out} \Delta d$ is the cross-section area of the junction wall and $E_{\rm wall}=100$MPa is elastic modulus of the cell wall \cite{tuson_measuring_2012}. Inserting the values gives $k\approx 1.2\mu$N/$\mu$m. Figures 3c,d show that the fraction $x$ of transconjugants is much higher than in Model 2 for low critical strains $\epsilon_0<0.1$. The stabilising effect of the junction has thus an important effect on the efficiency of conjugation. 

\begin{figure}
\includegraphics*[width=\columnwidth, bb=30 60 560 412]{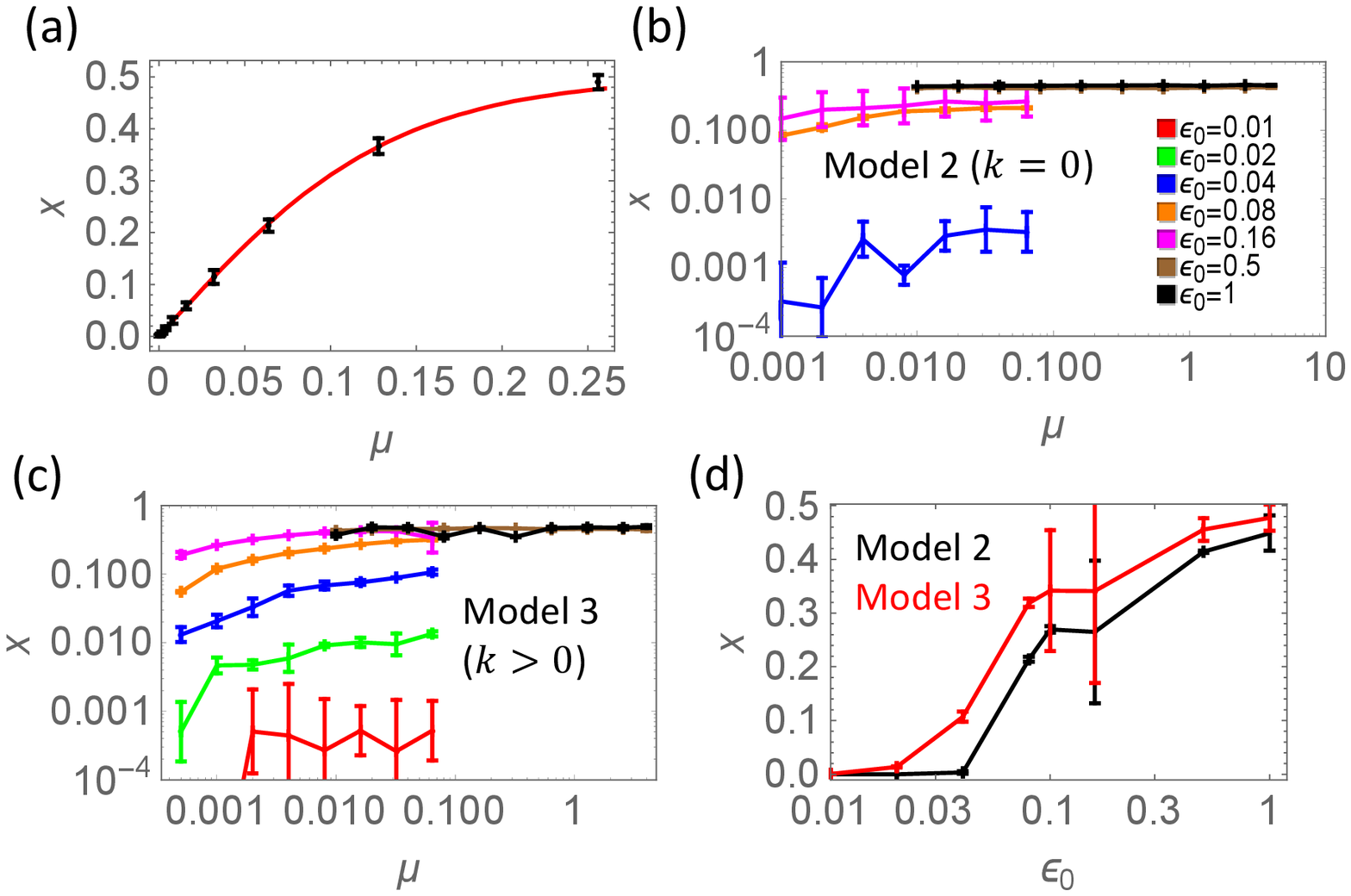}
\caption{Conjugation in exponentially expanding colonies for $N_0=250$ and $N=2000$. (a) Fraction $x$ of transconjugants as a function of the conjugation rate $\mu$ in Model 1. Points = computer simulations, red line = theoretical prediction from Eq. (\ref{eq:xwm}) with the best-fit value $q=7.0$. (b) $x$ versus $\mu$ for Model 2 ($k=0$), for different critical strains $\epsilon_0$. Data points for $\epsilon_0=0.01$ and $0.02$ are absent because no conjugation was detected for such low $\epsilon_0$.
(c) $x$ versus $\mu$ for Model 3 ($k=1.2\mu$N/$\mu$m). (d) $x$ versus $\epsilon_0$ for Models 2 and 3, for $\mu=0.016$h$^{-1}$. In all plots, errors are s.e.}
\end{figure}

{\it Colliding colonies.}
We have seen that the relative movement of cells reduces the rate of conjugation because conjugative junctions  break before plasmid transfer is completed. This suggest that conjugation should be extremely slow if donor and acceptor cells move rapidly towards each other. Such a  situation will occur in ``collisions'' of microbial colonies initiated from two cells separated by distance $b$ (Fig. 1d). As the two colonies grow, they eventually collide and merge (Fig. 3a).

We simulated this scenario using Model 3. We assume that nutrients are abundant and all cells replicate at the same rate; this is realistic for small inter-colony distances $b$ such as $b<40\mu$m considered here. The simulation is stopped when the length of the collision front reaches $L=25\mu$m, at which point we count the number of transconjugants. Figure 3b shows how the density of transconjugants at the front $n_{\rm T}/L$ varies with the initial distance $b$ for the parameters $k=1.2\mu$N$/\mu$m, $\mu=0.06{\rm h}^{-1}$, and a range of $\epsilon_0=0.025-0.5$. The density of transconjugants decreases with increasing $b$ for all $\epsilon_0$. This can be intuitively understood as a combination of two effects. First, colonies collide earlier for smaller $b$, thus giving bacteria more time to conjugate before the collision front reaches the length $L$ at which we count transconjugants. Second, cells move faster at the edge of larger colonies and hence remain in contact for a shorter time upon collision. An exponentially-growing colony will have $N\approx 2^{t/T}$ cells at time $t$, where $T$ is the average doubling time. The radius $R\propto \sqrt{N} = 2^{t/(2T)}$ and the velocity of the edge $v={\rm d}R/{\rm d}t \propto \ln(2)/(2T) 2^{t/(2T)}$, hence $v= R \ln(2)/(2T)$ is proportional to the colony's radius. The relative average speed of donors versus recipients near the front will be thus $v=(b/T)\ln(2)$ upon collision. Figure 3c shows that cells at the front do not stop moving: 
the relative velocity is comparable to that immediately before collision. This is not caused by inertia (cells in our model do not have inertia) but by cells resisting compression while the merged colony is expanding. The relative velocity of donors and recipients is proportional to the initial separation $b$; this  reduces the time of physical contact for larger $b$ and lowers conjugation efficiency.

To investigate the role of junction stiffness on conjugation efficiency, we varied $k$ while keeping other parameters fixed (Fig. 3d). The density of transconjugants $n_T/L$ initially increases with increasing $k$ and then levels off above $k>5\mu$N/$\mu$m if the critical strain is small. The plateau occurs for smaller $k$ if the critical strain is larger. This further stresses the importance of the stabilizing effect of the junction, but it also shows that making the junction very stiff does not pay off if this is not accompanied by an increase in the critical strain $\epsilon_0$.

\begin{figure}
\includegraphics*[width=\columnwidth, bb=40 65 560 450]{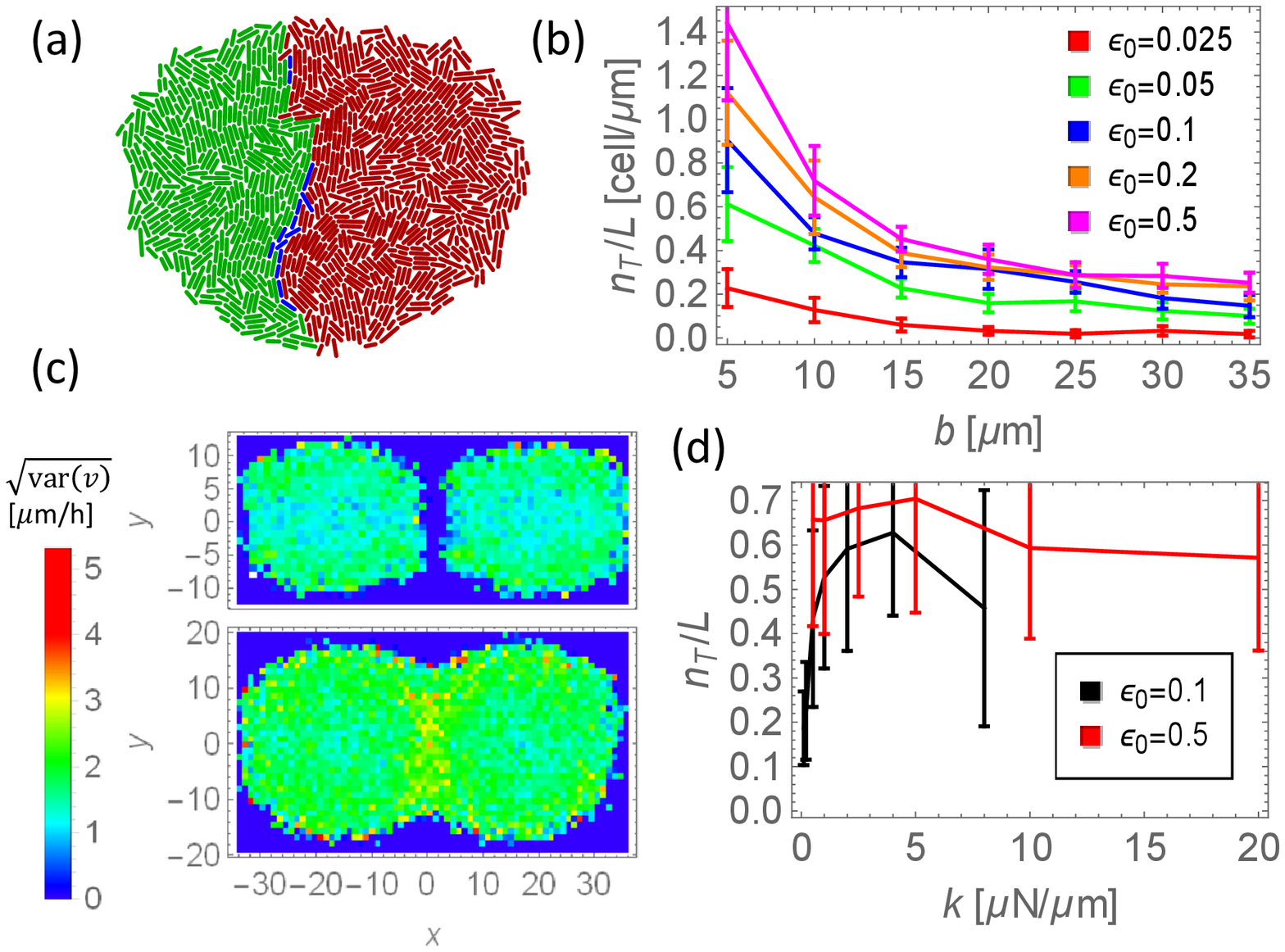}
\caption{Conjugation in colliding, exponentially growing colonies. (a) Example of two colonies colliding and partially merging. Red = donors, green = recipients, blue = transconjugants. (b) Density of transconjugants at the interface between the two colliding colonies plotted versus the initial distance $b$. (c) Density plot of the standard deviation $\sigma=\sqrt{{\rm var}(v)}$ of cell velocities in the colonies. $\sigma$ measures the difference in the speed of neighbouring cells. The upper plot corresponds to the time just before and the lower plot to the time just after the colonies merged. (d) Density of transconjugants as a function of spring constant $k$, for two different $\epsilon_0=0.08$ (black) and $0.32$ (red). Errors are s.e.}
\end{figure}

{\it Linearly-expanding colony.}
We now consider the same scenario as in Ref. \cite{hallatschek2007genetic} -- a radially expanding colony starting from a 1:1 mixture of donors and recipients. Simulations are initialized with $N_0=100$ bacteria on a ring of radius $R=15\mu$m (Fig. 1e) with randomly assigned types (donors, recipients). Cells consume nutrient as they grow. This causes the formation of a layer of replicating cells surrounding a static ``core'' following nutrient depletion in the colony centre (Fig. 4a). The simulation is stopped when the colony reaches $N=5000\dots5\times10^{5}$ cells. 
In the absence of conjugation the two types of cells segregate into sectors (Fig. 4b). This process is well understood (see e.g. \cite{hallatschek2007genetic,hallatschek2010life,korolev2010genetic,korolev2012selective}) and is driven by fluctuations in the number of cells in the active layer. This causes some sectors to grow at the expense of other sectors that collapse, and the dynamics of sectors can be mathematically described in terms of annihilating random walkers representing the sectors' boundaries \cite{hallatschek2010life}. The theory can also be extended to include non-identical growth rates (selection) \cite{korolev2012selective}, and cooperation between cells (mutualism) \cite{lavrentovich_asymmetric_2014}.

Here we focus on how conjugation and its associated physical forces affect segregation.
We first consider Model 1 and investigate how the number of sectors depends on the conjugation initiation rate $\mu$.   
Figure 4b shows simulation snapshots for low and high $\mu$. Even a relatively small ($6\times$) change in $\mu$ leads to a qualitatively different pattern, with many sectors in the low-$\mu$ case and only one sector (plus a few streaks of acceptor cells) in the high-$\mu$ case. This shows that the number of sectors $N_S$ may be a sensitive measure of conjugation efficiency. Moreover, $N_S$ is easy to determine in wet-lab experiments if plasmid-carrying cells have a distinct phenotype, e.g. they are fluorescent. In our simulations however, instead of counting the sectors directly, we calculate $N_S$ as
\bq
	N_{S}=\left(\sum_{i=1}^{n}\frac{\phi_{i}^{2}}{4\pi^{2}}\right)^{-1}, \label{sect_weight}
\eq
where $\phi_{i}$ is the angular size of sector $i$. When all sectors have the same size, $N_S$ is equal to the true number of sectors. 
If the sectors have unequal size, small sectors contribute less to $N_{S}$ than the large ones. Thus equation (\ref{sect_weight}) tends to smooth out $N_S$ and helps to reduce statistical uncertainties.

Figure 4c shows the number of sectors $N_S$ thus calculated, as a function of $\mu$, averaged over many simulated colonies. $N_S$ decreases with increasing $\mu$, until eventually only one sector remains for sufficiently large $\mu$. For smaller $\mu$, $N_S$ decreases with increasing size $N$ of the colony. In Appendix B we argue that when $N\to\infty$, only one plasmid-carrying sector survives for any $\mu>0$.

Let us now consider Models 2 and 3. 
Figure 4d shows the number of sectors for fixed $\mu=0.06$h$^{-1}$, different final colony sizes $N$, and two different junction spring constants: $k=0$ and $k=1.2\mu$N/$\mu$m. There are many more sectors in the case in which the junction does not exert force ($k=0$). When $k>0$, the junction stabilizes the mating pair against relative displacements of cells. This increases the effective rate of conjugation and decreases the number of recipient sectors. 

\begin{figure}
\includegraphics*[width=\columnwidth, bb=30 60 565 437]{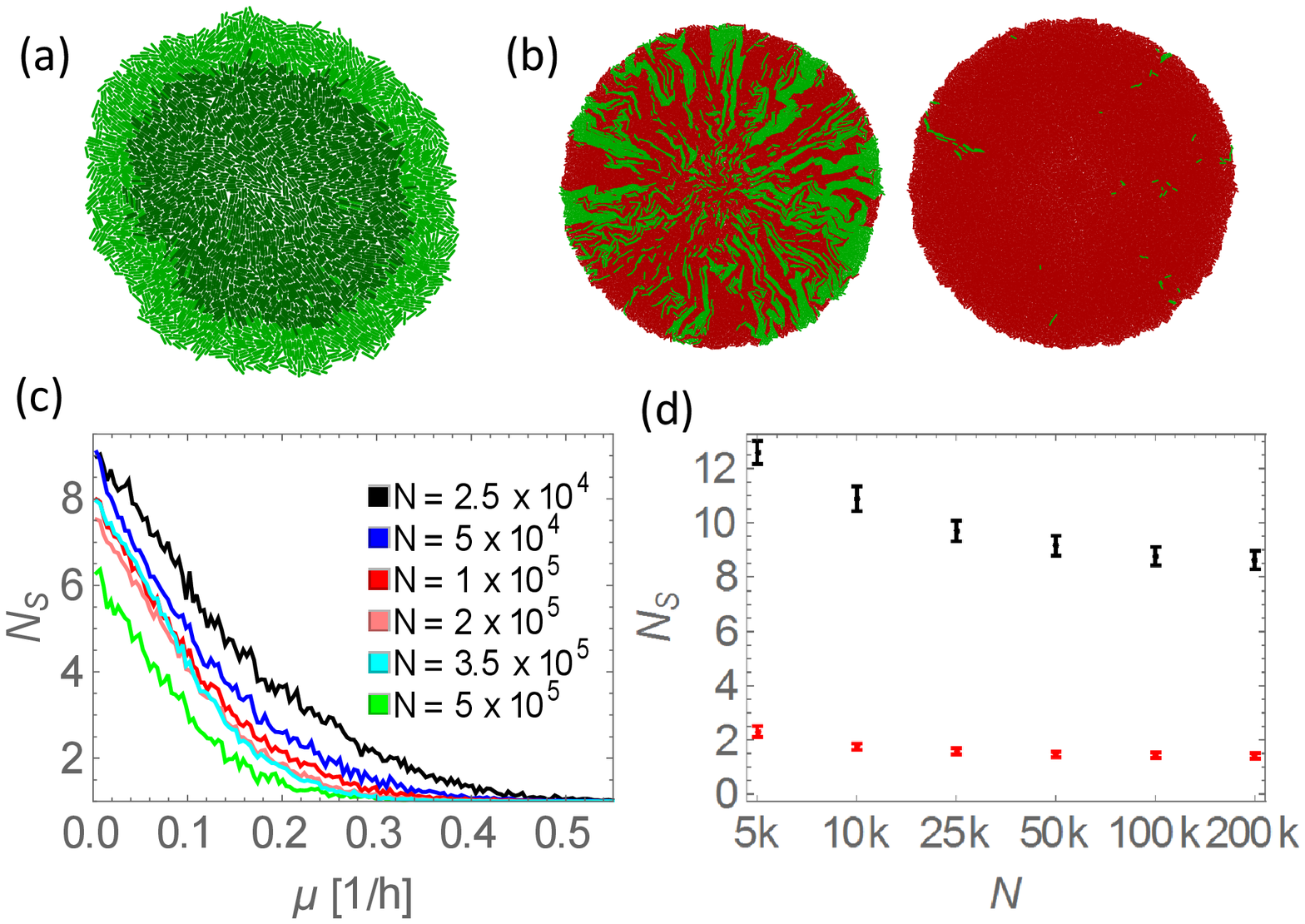}
\caption{Conjugation in linearly expanding colonies. (a) A simulation snapshot showing the growing layer (light green) and the static core (dark green) in the centre where nutrients have been depleted. (b) Sectors (green = recipients, red = donors and transconjugants) for two conjugation initiation rates $\mu=0.05$h$^{-1}$ (left) and $\mu=0.3$h$^{-1}$ (right). In both cases, the total number of cells $N=16000$. (c) Average number of sectors $N_S$ versus $\mu$ in Model 1, for different final sizes $N$ of the colony. (d) $N_S$ versus $N$ for Model 2 ($k=0$, black) and Model 3 ($k=1.2\mu$N/$\mu$m, red). Errors are s.e. }
\end{figure}

{\it Conclusions.}
In this work we have used a computer model to investigate conjugation in bacterial colonies. Compared to previous work that studied the effect of nutrient concentration \cite{fox_spatial_2008,merkey_growth_2011}, spatial distribution of cells \cite{krone_modelling_2007,freese_genetic_2014}, or the fitness cost of the plasmid on the dynamics of conjugation \cite{krone_modelling_2007,lenski_bacterial_2010}, our model includes explicit physical interactions: cell-cell repulsion and cell-cell attraction caused by the formation of the conjugative junction. We have shown that the relative movement of donors and acceptors significantly affects the observed rate of conjugation, even if conjugation is initiated with high frequency. This is because conjugation takes a finite time to complete, and if two conjugating cells move away from each other during this time, conjugation will be interrupted. This process can be mitigated if the conjugative junction is able to exert force that counteracts the movement of the two cells. 

Computer models and experiments show that mechanical interactions between bacterial cells are an important aspect of bacterial life. Mechanical interactions are responsible for nematic ordering in colonies confined in microfluidic devices \cite{volfson2008biomechanical}, different shapes of colonies growing on agarose \cite{farrell_mechanically_2013,giverso_branching_2015}, and the magnitude of genetic drift in biological evolution experiments \cite{farrell_mechanical_2017}. Our work provides another example of the role of mechanical interactions in bacterial colonies and biofilms.

Our results may help to explain some experimental findings. For example, conjugation occurs with low yield when recipient and donor colonies collide \cite{reisner_situ_2012} as predicted by our model. Shear forces created by growth could also explain low efficiency of conjugation in some biofilms \cite{christensen_establishment_1998,krol_increased_2011}. The stabilising role of the conjugative junction has been suggested to play a certain role in biofilms \cite{ghigo2001natural}; here we show why this should be the case.

Quantitative predictions of our models could be experimentally validated using e.g. the colliding colony assay. 
The experiment could be set up as follows. Recipient cells are made to constitutively express a fluorescent protein (say a cyan fluorescent protein, CFP) from a gene on the chromosome. Donor cells are also made fluorescent by putting an appropriate gene (coding for, say, a yellow fluorescent protein, YFP) on the plasmid.
A dilute mixture of recipient and donor cells is spread on a thin agarose pad, allowed to dry, and covered with a microscopic slide. Growing colonies can be imaged using fluorescence microscopy and the number of transconjugants assessed by counting the number of cells that are both cyan- and yellow-fluorescent. The initial separation can be accurately measured at the start of the experiment for many pairs of donor-acceptor cells on the same agarose pad. If our theory is correct, the density of transconjugants should decrease with increasing initial separation $b$ of the colonies. 

Our results have implications for biological evolution. If conjugation is affected by mechanical forces, it should occur more often in situations in which shear forces acting on the mating pair are weak. Interestingly, sessile communities of microorganisms such as biofilms may be subject to larger shear forces than mating pairs in liquids because growth can generate large forces (nN in our simulations). Analogous forces acting on a pair of cells in a suspension will be much smaller in a typical situation which bacteria encounter in nature (e.g. in a river or the animal gut). We can estimate the required fluid velocity gradient $\Delta u/\Delta z$ from the Stokes formula: $F=6\pi \eta R (\Delta u/\Delta z) l$ where $R$ is the size of the cell, $l$ is the separation between two mating cells, and $\eta$ is the dynamic viscosity of the fluid. Inserting $F=10$nN, $\eta=10^{-3}$Pa$\cdot$s, and $R=l=2\mu$m, we obtain that $(\Delta u/\Delta z)$ would have to be $1.3\times 10^5$(m/s)/m - clearly a very high shear rate not expected in most natural situations, except perhaps close to surfaces. Conjugation may therefore occur more easily in dense bacterial dispersions than in biofilms.

Our models could also be used to infer the apparent conjugation rate from the experimentally determined number of sectors in co-culture experiments. This could be a simpler alternative to the existing methods used for experimental measurements of conjugation rates in spatially structured populations, which often require counting different types of cells. Counting sectors in macroscopic colonies of $\sim$cm size is much less time consuming: it does not require sophisticated equipment (microscopes) or plating out many replicates to obtain good statistics. If a fluorescent plasmid is used, the experiment can be performed with a simple illumination plate and a suitable optical filter.

{\it Acknowledgements.} We thank R. Allen (Edinburgh) and B. Smets (TU Denmark) for discussion. JP was supported by an EPSRC studentship. BW was supported by a Royal Society of Edinburgh/Scottish Government Personal Research Fellowship.

\bibliographystyle{apsrev-nourl}
\bibliography{literature}

\section{Appendix A - details of computer simulations}
Nutrient dynamics is simulated using the diffusion equation with sinks:
\bq
	\frac{\partial c}{\partial t} = D \left( \frac{\partial^2 c}{\partial x^2} + \frac{\partial^2 c}{\partial y^2} \right) - \sum_i u_i \delta\left(\vec{r}_i - \vec{r}\right)
\eq
Here $\vec{r}_i=(x_i,y_i)$ is the position of $i$-th cell, $c=c(\vec{r},t)$ is the nutrient concentration at position $\vec{r}$ and time $t$, $D$ is the diffusion coefficient of the nutrient, and $u_i$ is the nutrient uptake rate. Nutrients are initially distributed uniformly and their concentration is $c(\vec{r},0)=c_0=1$. 
Bacteria take up nutrients at rate $u=Af(c)$, where $A=\pi a^2/4+la$ is the assumed nutrient-absorbing surface area of the cell and $f(c)=c/(c+c_{1/2})$ is the Monod function with a half-saturation constant $c_{1/2}$ taken to be $1$ for simplicity. 
We also assume $f(c)=0$ for $c<c_{\rm crit}$ where $c_{\rm crit}=0.2$ is the minimum nutrient concentration below which growth ceases. 

The elongation rate of the bacterial cell is ${\rm d}l/{\rm d}t=gAf(c)$ and the proportionality coefficient $g$ is such that if nutrient is abundant ($c_0\gg c_{1/2}$), cells divide every $25$min. Upon reaching the maximum length $l_{\rm max}$ drawn from the normal distribution with mean $4\mu$m and standard deviation $0.6\mu$m a cell divides into two daughter cells; each cell is rotated by a small random angle $\Delta\phi\in (-0.001,0.001)$ to prevent (unrealistic) perfect alignment. The length and the growth rate are based on typical values for rod-shaped bacteria such as {\it E. coli} growing on rich media \cite{farrell_mechanical_2017}.

The position $\vec{r}_i$ of the centre of mass and the angular coordinate $\phi_i$ of the major axis of cell $i$ evolve according to Newton's overdamped equations of motion:
\ba
	\frac{d\vec{r}_i}{dt} = \vec{F}/(\zeta m), \label{eq:m1} \\
	\frac{d\phi_i}{dt} = \tau/(\zeta J). \label{eq:m2}
\ea
Here $\vec{F}$ and $\tau$ are the total force and torque acting on the cell, $m=\pi (a^2/4)(l+2a/3)$ is the mass (pg if $a,l$ are in $\mu$m), $J=\pi (a^2/4) ((3/16)la^2+(1/15)a^3+(1/6)l^2a+(1/12)l^3)$ is the momentum of inertia, and $\zeta=2.6\times 10^{18}$h$^{-1}$ is the damping coefficient. The above $\zeta$ gives the friction force of $10$nN for a cell of mass $5$pg moving with speed $10\mu$m/h \cite{farrell_mechanically_2013}. Equations (\ref{eq:m1}-\ref{eq:m2}) are integrated using the Euler method with a time step $\Delta t=2^{-14}$h.

\section{Appendix B - conjugation in linearly expanding colonies}
We can show that, for sufficiently large $N$, the plasmid is always expected to invade the whole population so that only one sector remains, even if the plasmid does not convey any growth advantage. 
Assume the colony is well in the linear phase of growth (radius $R\propto t$). The angular size $\phi$ of a transconjugant sector obeys the following equation
\bq
	\frac{{\rm d}\phi}{{\rm d}R} = \frac{\nu}{R} + \frac{\eta(r)D_{\rm sector}}{R}.  \label{eq:phi}
\eq
The equation has two terms. The term $\nu/R$ corresponds to the expansion of the sector due to conjugation occurring at its boundaries; the parameter $\nu$ is proportional to the conjugation rate $\mu$. The term $\eta(r)D_{\rm sector}/R$ corresponds to Brownian diffusion of the sector boundaries \cite{hallatschek2010life}; $D_{\rm sector}$ is the effective wall diffusion constant. If the sector does not collapse due to a random fluctuation, the average $\phi$ calculated from Eq. (\ref{eq:phi}) is given by
\bq
	\left<\phi\right> \cong \mu \ln(R/R_0),
\eq
where $R_0$ is the initial radius of the colony. Although probability that a sector collapses is non-zero, the initial number of sectors (equal to the initial number of donor cells) is large enough to ensure that at least one donor sector survives and continues to grow.
Thus for $R\to\infty$ only one sector encompassing the entire active layer remains. However, since $\left<\phi\right>$ grows only logarithmically with the radius of the colony, a finite colony will typically have a finite number of sectors unless $\mu$ is very large.

\end{document}